\documentclass[twocolumn,aps,floats,prd,amsmath,amssymb]{revtex4b3}
\begin{document}

\draft \tolerance = 1000

\setcounter{topnumber}{1}
\renewcommand{\topfraction}{1}
\renewcommand{\textfraction}{0}
\renewcommand{\floatpagefraction}{1}
\newcommand{\br}{{\bf r}}

\twocolumn[\hsize\textwidth\columnwidth\hsize\csname@twocolumnfalse\endcsname

\title{{\rule{17cm}{.15mm}{\bfseries{\\\vspace{3mm}Standard Cosmology Through Similarity\\\rule{17cm}{.15mm}}}}}
\author{Jos\'e Antonio Belinch\'on}
\address{Grupo Inter-Universitario de An\'alisis Dimensional\\
Dept. F\'isica ETS Arquitectura  UPM\\
Av. Juan de Herrera 4 Madrid 28040 Espa\~na\\
E-mail: abelinchon@caminos.recol.es}
\date{October 1999}

\begin{abstract}
\emph{~~~~In this paper we apply dimensional analysis (D.A.) to two cosmological
models: Einstein-de Sitter and one Friedmann-Robertson-Walker (FRW) with
radiation predominance. We believe that this method leads to the simplest
form of solution the differential equations that arise in both models and
would be useful as a base for the solution of more complex models. The aim
of the paper is therefore rather pedagogical since it tries to show
different dimensional techniques.}\medskip

\end{abstract}\vspace{.4cm}
\maketitle
]
\section{\textbf{Introduction.}}

In this paper we try to show how Dimensional Analysis (D.A.) can be formally
applied to cosmology. With this aim we shall study two models of known
solutions: the Einstein-de Sitter, and one model type
Friedmann-Robertson-Walker (FRW) with radiation predominance (see \cite{N}),
we shall prove, that we can reach the known results and consequently the
usefulness of D.A. as a tool to solve this type of models and eventually
those founded on some difficult equations like, for example, alternative
theories of the gravitation that envisage the ``constants'' as scalars
functions dependent on time $t$ (see \cite{T}).\bigskip 

Now, we go next to explain step by step the followed dimensional method to
obtain a complete solution to the equations that govern each one of the
models. We begin in the second section with a brief account of the
fundamentals of D.A. Although a more complete understanding is to be found
through the advised literature (see \cite{P} and \cite{B}), we hope that
such account might suffice for the practical application of D.A. We
accompany the explanation with a short amount of examples (that reduce to a
single one) for a better understanding. We shall follow a simple scheme.
Firstly we shall calculate the multiplicity of the dimensional base to be
used for the model under study, choosing one of the possible bases. In
second place we shall select the set of quantities and constants to consider
in each model since, as stated in third section we are only interested in
the fundamental quantities and the unavoidable constants. Then with the help
of the Pi theorem we will arrive to the solution of the governing equations
of our two models. In the third section the Einstein-de Sitter model will be
studied and in the fourth a model type FRW with radiation predominance from
our dimensional perspective. In last section the solution is found in
function of two monomials related through an unknown function, we will take
into account the Barenblatt`s criterium (see \cite{B}) to obtain a complete
solution to the model. We will justify, from a dimensional point of view,
the utilization of the Planck system of units (see \cite{L}), making use of
this criterion. In each case we shall contemplate an alternative way to
avoid the Barenblatt criterium and solve it completely the problem.

\section{\textbf{Dimensional analysis}.}

In our opinion J. Palacios (see \cite{P}) has been the first author that has
tried to arise Dimensional Analysis from a ``Method'' to a ``Theory'', where
all of its results are derived from a very limited number of postulates
(two). Although the theory of Palacios was in our view mostly successful a
long time has elapsed and his postulates have been amended by M. Casta\~{n}s
(see \cite{C}) According to Palacios it is possible to select the universal
and unavoidable constants as relating two inseparable quantities, ``that is,
two quantities such that the presence of one of them in a body involves the
presence of the other in the same body, so that equal amounts of the first
one compounds to equal amounts of the second'' one. Palacios, then,
postulates ($2^{nd}$ postulate): ``unavoidable universal constants are those
which relate to inseparable quantities all others are superfluous''. The
corrections to this postulate by M. Casta\~{n}s are too slight to be
included here as a consequence of the corrected postulate and on account
that the constants $c,\epsilon _{0},\mu _{0}$ are connected by a well known
relation in such a way that the remaining ones in Physics are only five
independent constants $G,c,\hbar ,k_{B}$ and either $\epsilon _{0}$ or $\mu
_{0}.$ It is to be observed that $G,c,\hbar ,k_{B}$ are the constants
selected by Planck to establish his system of absolute units.\bigskip 

The first postulate reads:

\emph{``The fundamental laws can be chosen in such a way that they are
relations of proportionality of defined powers of the quantities involved ''.%
}\ \bigskip

It is, in the opinion of M. Casta\~{n}s, advantageously replaced by the
following statement:\bigskip

\emph{`` Dimensional Analysis can be applied to those Physical Theories
whose fundamental laws may be written:} 
\begin{equation}
f\text{(}\pi _{1}\text{,.....,}\pi _{n}\text{)=0}
\end{equation}
\emph{where $\pi _{i}$ they are dimensionless products ($\pi -$%
monomials)''}.\bigskip

Consequences from both postulates (or rather of the second one and the
quoted statement) are:

\begin{enumerate}
\item  It is possible to fix the multiplicity of the dimensional base.

\item  The $\pi $ (Buckingham) theorem may be proved and enunciated without
ambiguity.
\end{enumerate}

The recipe to fix the multiplicity of the base is:\bigskip

\emph{\ ``The number of quantities forming the dimensional base (its
multiplicity) is given by the difference }$m=n-h$\emph{\ between the total
number of quantities (including unavoidable constant) and the rank of the
matrix formed with the exponent occurring in the monomial present in the
fundamental equations of the concerned theory .''\bigskip }

\emph{{\LARGE E}xample}:\textbf{\ Calculation of multiplicity of dimensional base: }The fundamental
equation of thermal radiation is Planck law: 
\begin{equation}
\pi _{1}=(e^{\pi _{2}}-1)^{-1}
\end{equation}
where 
\begin{equation}
\pi _{1}=\frac{u_{\gamma }c^{3}}{8\pi h\gamma ^{3}}\qquad \qquad \pi _{2}=%
\frac{h\gamma }{k_{B}T}
\end{equation}
Obviously the first postulate is not fulfilled. Instead we use the Casta\~{n}%
s statement. The matrix is: 
\begin{equation}
\begin{array}{r|rrrrrr}
& u_{\gamma } & c & h & \gamma  & k_{B} & T \\ \hline
\pi _{1} & 1 & 3 & -1 & -3 & 0 & 0 \\ 
\pi _{2} & 0 & 0 & 1 & 1 & -1 & -1
\end{array}
\end{equation}
with rank $h=2$.

Then, the multiplicity is $m=n-h=6-2=4$ and we can take a dimensional base
of four quantities, for example,$\left\{ L,M,T,\Theta \right\} $, where $%
\Theta $ means '' dimensions of temperature ''. Also a base can be formed by 
$[\gamma ]=T^{-1}$ and the constants $c,h,k_{B}$ . However the same results
will be obtained by making use of a base of higher multiplicity (f.e. $%
\left\{ \gamma ,c,h,k_{B},M\right\} $).\bigskip 
\bigskip

The number of independent monomials (dimensionless products) is $i=n-j$
where $j$ is the rank of the matrix of the dimensional exponents of the
quantities (and unavoidable constants) relative to a suitable base.\bigskip

\emph{{\LARGE E}xample}:\textbf{A Significant Example:Planck System. }As already mentioned, the
system of absolute units proposed by Planck consists of the four universal
constants $\left\{ G,c,h,k_{B}.\right\} :$%
\begin{equation}
\begin{array}{r|rrrr}
& l_{P} & G & c & \hbar  \\ \hline
L & 1 & 3 & 1 & 2 \\ 
M & 0 & -1 & 0 & 1 \\ 
T & 0 & -2 & -1 & -1
\end{array}
\Longrightarrow \pi _{1}=\frac{G\hbar }{l_{p}^{2}c^{3}}
\end{equation}
\begin{equation}
\begin{array}{r|rrrr}
& m_{P} & G & c & \hbar  \\ \hline
L & 0 & 3 & 1 & 2 \\ 
M & 1 & -1 & 0 & 1 \\ 
T & 0 & -2 & -1 & -1
\end{array}
\Longrightarrow \pi _{2}=\frac{\hbar c}{m_{p}^{2}G}
\end{equation}
\begin{equation}
\begin{array}{r|rrrr}
& t_{P} & G & c & \hbar  \\ \hline
L & 0 & 3 & 1 & 2 \\ 
M & 0 & -1 & 0 & 1 \\ 
T & 1 & -2 & -1 & -1
\end{array}
\Longrightarrow \pi _{3}=\frac{G\hbar }{t_{p}^{2}c^{5}}
\end{equation}
etc... In the Matrix it is shown how to obtain the mechanical quantities of
length, time and mass by using a base $L,T,M.$ We get the monomials $\pi _{1}
$, $\pi _{2}$ and $\pi _{3}$. Then: 
\begin{equation}
l_{p}=\sqrt{\frac{G\hbar }{c^{3}}}\qquad t_{p}=\sqrt{\frac{G\hbar }{c^{5}}}%
\qquad m_{p}=\sqrt{\frac{\hbar c}{G}}
\end{equation}

To obtain Planck`s temperature we should consider that its product by $k_{B}$
has dimensions of energy. This is equivalent to use a dimensional base that
includes temperature.

\section{\textbf{Einstein-de Sitter.}}

In this section we shall study the solution of the Einstein-de Sitter model
by D.A. We begin showing the essential features of this cosmological model,
then we pass to present the field equations that describe it. Departing of
these and continuing the traced plan in the previous section we will
calculate the multiplicity of the dimensional base and we will choose one of
the possible ones. We shall classify the set of quantities and constants and
end resolving the equations through dimensional analysis, that is to say,
through the Pi theorem.\bigskip

Our three ingredients of relativistic cosmology are as follows: (we use the
standard notation).

\begin{enumerate}
\item  The cosmological principle which leads to the Robertson-Walker line
element, 
\begin{equation}
ds^{2}=-c^{2}dt^{2}+f^{2}(t)\left[ \frac{dr^{2}}{1-kr^{2}}+r^{2}d\theta
^{2}+r^{2}\sin {}^{2}\theta d\phi ^{2}\right]   \label{olg1}
\end{equation}

\item  Weyl `s postulate which requires that the substratum is a perfect
fluid

\begin{equation}
T_{ij}=(\rho +p)u_{i}u_{j}-pg_{ij}  \label{olg2}
\end{equation}

\item  General relativity
\end{enumerate}

\begin{equation}
R_{ij}-\frac{1}{2}g_{ij}R=\frac{8\pi G}{c^{4}}T_{ij}\qquad \qquad
div(T_{i}^{j})=0\qquad   \label{f9}
\end{equation}

The model is constituted by a perfect fluid with $\rho _{m},$ $k=$ $p=0.$
where $\rho _{m}$ represents the matter density of all galaxies. Under these
considerations the field equations are as follows:

\begin{equation}
\begin{array}{l}
\left( 12.1\right) \qquad 2ff^{\prime \prime }+\left( f^{\prime }\right)
^{2}=0 \\ 
\left( 12.2\right) \quad \quad 3\left( f^{\prime }\right) ^{2}=8\pi G\rho
_{m}f^{2}\quad \quad  \\ 
\left( 12.3\right) \qquad f\rho _{m}^{\prime }+3\rho _{m}f^{\prime }=0\quad
\Longrightarrow \rho _{m}=M\,f\,^{-3}
\end{array}
\label{e1}
\end{equation}
equation $4.3$ is also considered as state equation.

\subsection{\textbf{Multiplicity of the dimensional base.}}

We now calculate the multiplicity of the dimensional base. For this purpose
we observe that equation (\ref{e1}) can be written (see \cite{P} and \cite{C}%
): 
\begin{equation}
\begin{array}{l}
\left( 12.1\right) 2\frac{f^{\prime \prime }}{f}+\left( \frac{f^{\prime
}}{f}\right) ^{2}=0~ f^{\prime }=\left[ \frac{df}{dt}\right] =\frac{f}{t}%
\;f^{\prime \prime}=\left[ \frac{d^{2}f}{dt^{2}}\right] =\frac{f}{t^{2}}
\\ 
\left( 12.2\right) \quad 3\left( \frac{f^{\prime }}{f}\right) ^{2}=8\pi G\rho
_{m}\quad \quad \left[ \left( \frac{f^{\prime }}{f}\right) ^{2}\right]
=t^{-2} \\ 
\left( 12.3\right) \quad \rho ^{\prime }+3\rho \frac{f^{\prime }}{f}=0\qquad
\Longrightarrow \rho _{m}=\frac{M\,}{f\,^{3}}\text{ }\qquad 
\end{array}
\end{equation}
These equations can be written in a dimensionally equivalent form 
\begin{equation}
\begin{array}{l}
(12.1)\quad \frac{{f}}{t^{2}}\frac{1}{{f}}+\frac{1}{t^{2}}=0 \\ 
(12.2)\quad \frac{1}{t^{2}}=G\rho _{m}\ \qquad \Rightarrow G^{-1}\rho
_{m}^{-1}t^{-2}=1 \\ 
(12.3)\quad \rho _{m}=\frac{M\,}{f\,^{3}}\ \qquad \Rightarrow \text{ }\rho
_{m}f^{3}M^{-1}=1
\end{array}
\end{equation}
that leads to the following dimensionless products (see \cite{P} and \cite{C}%
) 
\begin{equation}
\Longrightarrow \left\{ 
\begin{array}{l}
\pi _{1}:=\text{ }t^{-2\,}G^{-1}\,\rho _{m}^{-1} \\ 
\pi _{2}:=\text{ }\rho _{m\,}\,M^{-1}\,f\,^{\,3}
\end{array}
\right. 
\end{equation}
from the first equation $(4.1)$, we do not obtain dimensional information.
We proceed to calculate the multiplicity of the base of this model. The
range of the matrix of the exponents of the quantities and constants
included in the monomials is $2$ as it results immediately from: 
\begin{equation}
\begin{array}{r|rrrrr}
& \rho _{m} & G & f & t & M \\ \hline
\pi _{1} & -1 & -1 & 0 & -2 & 0 \\ 
\pi _{2} & 1 & 0 & 3 & 0 & -1
\end{array}
\end{equation}
The multiplicity of the dimensional base is therefore $m=(number$ $of$ $%
quantities$ $and$ $constants)-(range$ $of$ $the$ $matrix)$, in this case it
is $m=3.$ Thus we can use the base of classical mechanics $B=\left\{
f,M,t\right\} \approx \left\{ L,M,T\right\} .$ Other base could be $%
B^{\prime }=\left\{ \rho ,G,t\right\} .\bigskip $

\textbf{Remark}The constant $c$ does not appear in the equations (\ref{e1}). This justifies
the utilization of Newton`s mechanics in the study of the model and
simplifies the solution of such equations, that it is reduced to a single $%
\pi -$monomial.

\subsection{\textbf{Quantities and constants.}}

In the case of geometric models, like the previous one, it is trivial the
election of the fundamental quantities, since it can be mathematically
demonstrated, making use of the Killing`s equations that $\left\{ t\right\} $
is our fundamental quantity. In the case of problems with more physical
content we must appeal to our physical knowledge of the problem to be able
to choose the set of fundamental quantities (or governing parameters in the
nomenclature of Barenblatt). The application of a FRW metric, implies that
our universe will be homogeneous and isotropic. This also implies, as it can
be proved, that all the quantities that appear in the equations should be
functions only on $\left\{ t\right\} .$ That is to say; the radius of the
universe $f,$the energy (matter) density $\rho $ and the expansion speed $v$
are functions on $\left\{ t\right\} $ and of the unavoidable constants.
Therefore, we will say that $f$ ,$\rho $ and $v$ are derived quantifies
whereas $t$ is the single fundamental quantity that appears in the model.
The physical and characteristic constants are $G$ and $M$ respectively. The
rest of the quantities of the model depends on $t$ and on the set of
constants that appear in the equations of the model, in this case $\left\{
G,M\right\} .$ The dimensional base of the model is $B=\left\{ L,M,T\right\}
,$ and the dimensions of the fundamental quantities and constants are:

\begin{enumerate}
\item  $t$ \emph{cosmic time (fundamental quantity) $\left[ t\right] =T$}

\item  \emph{$G$ Gravitational constant $\left[ G\right] =L^{3}M^{-1}T^{-2}$}

\item  \emph{$M$ The total mass of the Universe (characteristic constant of
the model). $\left[ M\right] =M$}
\end{enumerate}

The dimensional method that we follow consists therefore in: To calculate
the multiplicity of the dimensional base, to choose one of the possible
bases and to calculate the dimensional equations of each quantity with
respect to the elected base. Thus it is needed to make use of our physical
and mathematical knowledge of the problem in order to be able to choose the
fundamental quantities or governing parameters (in these models the only
fundamental quantity that appears is $t$, but in more complex models, also
more fundamental quantities or/and unavoidable universal and characteristic
constants might be considered). With these distinctions we shall calculate
the derived quantities ($f,\rho ,v$ etc...) through the Pi-theorem.

\subsection{\textbf{Solution of the equations through D.A.}}

Let us calculate through application of the Pi theorem, the radius of the
Universe $f(t)$, the velocity of expansion of the galaxies $v(t)$ and the
matter density $\rho (t)$ that contains the Universe with radius $f(t)$.
Therefore, we shall calculate these quantities in function of $\left\{
G,M,t\right\} ,$ that is to say, in function of the unavoidable constants $%
(G,M)$ and the fundamental quantity, $t,$ by making use of the dimensional
base $B=(L,M,T)$

We apply the Pi theorem to obtain:

\subsubsection{\textbf{Calculation of the radius of the Universe}}

The quantities that we will consider are: $f(t)\propto f$ $(G,M,t)$ in the
base $B=(L,M,T)$ where the dimensional equation of this quantity is $\left[ f%
\right] =L$ with respect to the base $B,$ we get: 
\begin{equation}
\begin{array}{r|rrrr}
& f & G & M & t \\ \hline
L & 1 & 3 & 0 & 0 \\ 
M & 0 & -1 & 1 & 0 \\ 
T & 0 & -2 & 0 & 1
\end{array}
\Rightarrow \pi _{3}=\frac{(GM)^{\frac{1}{3}}t^{\frac{2}{3}}}{f(t)}
\end{equation}
we have obtained a single monomial that brings us to the following solution: 
\begin{equation}
f(t)\propto (GM)^{\frac{1}{3}}t^{\frac{2}{3}}
\end{equation}
obtaining the result $f(t)\propto t^{\frac{2}{3}}$. Of course D.A. can not
find the value of the numerical and dimensionless constant of
proportionality.\bigskip

We can observe that the differential equation to be solved is: 
\begin{equation}
\left( \frac{f^{\prime }}{f}\right) ^{2}=\frac{8\pi G}{3}\frac{M}{f^{3}}%
\qquad \longmapsto \qquad f(f^{\prime })^{2}=\frac{8\pi GM}{3}  \label{ode1}
\end{equation}
that is immediately integrated after finding an adequate change of variable.
Our purpose is to show how the simplest dimensional technique solves
immediately the equation. We can explore other possibilities, since if we
observe the differential equation (\ref{ode1}) we can see that the constants 
$GM$ always keep the same relation into the equation. If we define a new
constant from them $K=GM$ where $\left[ K\right] =L^{3}T^{-2}$, the quantity 
$f$ \ can be recalculated but with respect to the next set of fundamental
quantities or governing parameters $M=M\left\{ t,K\right\} $ and $B=\left\{
L,T\right\} $%
\begin{equation}
\begin{array}{r|rrr}
& f & K & t \\ \hline
L & 1 & 3 & 0 \\ 
T & 0 & -2 & 1
\end{array}
\end{equation}
obtaining a single monomial that brings us again to the above solution. 
\begin{equation}
f\propto \left( Kt^{2}\right) ^{1/3}\Longrightarrow f(t)\propto (GM)^{\frac{1%
}{3}}t^{\frac{2}{3}}
\end{equation}

\subsubsection{\textbf{Calculation of the matter density.}}

The same discussion as above: $\rho _{m}(t)\propto \rho _{m}(G,M,t).$ where $%
\left[ \rho _{m}\right] =ML^{-3}$ 
\begin{equation}
\begin{array}{r|rrrr}
& \rho _{m} & G & M & t \\ \hline
L & -3 & 3 & 0 & 0 \\ 
M & 1 & -1 & 1 & 0 \\ 
T & 0 & -2 & 0 & 1
\end{array}
\Longrightarrow \pi _{4}=\frac{1}{\rho _{m}(t)Gt^{2}}
\end{equation}
this single monomial brings us to the following solution: 
\begin{equation}
\rho _{m}(t)\propto \frac{1}{Gt^{2}}
\end{equation}
It is somehow surprising that the result does not depend on $M$

\subsubsection{\textbf{Calculation of the velocity of expansion.}}

$v(t)\propto v(G,M,t)$ where $\left[ v\right] =LT^{-1}$%
\begin{equation}
\begin{array}{r|rrrr}
& v & G & M & t \\ \hline
L & 1 & 3 & 0 & 0 \\ 
M & 0 & -1 & 1 & 0 \\ 
T & -1 & -2 & 0 & 1
\end{array}
\Rightarrow \quad \pi _{5}=\frac{(GM)^{\frac{1}{3}}}{v(t)t^{\frac{1}{3}}}
\end{equation}
\begin{equation}
v(t)\propto (GM)^{\frac{1}{3}}t^{-\frac{1}{3}}
\end{equation}
In this model the application of the Pi theorem has carried us to obtain a
single dimensionless $\pi -$monomial, for each derived quantity.

\section{\textbf{FRW with radiation predominance.}}

We begin this section continuing the traced plan in section $2$, by
considering the equations that govern the model. These equations, like the
ones of the previous model, are based on the three exposed above basic
ingredients, equations $(\ref{olg1},\ref{olg2},\ref{f9})$. We shall select
the set of quantities and constants, solving the equations with the help of
the Pi theorem.\bigskip 

We will see that, in this case, the solution to be obtained for each one of
the calculated quantities depends on certain unknown function $\varphi ;$ $%
\pi _{1}=\varphi \left( \pi _{2}\right) $. To avoid this drawback we will
make use of the criterion of Barenblatt. This criterion will eventually
enable us to simplify the solution to one of type $\pi _{1}=\left( \pi
_{2}\right) ^{n}$ being possible to calculate through numerical methods the
appropriate value of $n$. In an alternative way, we shall solve again this
model by making use of the Planck system of units simplifying, from a
dimensional point of view, the solution obtained through the Baremblatt
criterion. We shall end showing how to avoid the use of Barenblatt criterium
by considering carefully the differential equations that govern the model.
We shall see how to reduce the number of constants and therefore the number
of $\pi -monomials$ in such a way that we shall obtain a single one which
keeps the problem perfectly solved.\bigskip

In this second case, a universe with radiation predominance and $\left(
k=0\right) $, the equations of Friedmann remain as follows: we maintain the
constants to show the dimensional wealth of the equations, 
\begin{equation}
\left\{ 
\begin{array}{l}
\left( 26.1\right) \qquad c^{2}\left( 2ff^{\prime \prime }+\left( f^{\prime
}\right) ^{2}\right) =-8\pi Gpf^{2} \\ 
\left( 26.2\right) \qquad 3c^{2}\left( f^{\prime }\right) ^{2}=8\pi G\rho
_{R}f^{2} \\ 
\left( 26.3\right) \qquad f\rho _{R}^{\prime }+3(p+\rho _{R})f^{\prime
}=0\Longrightarrow \rho _{R}f^{4}=A \\ 
\left( 26.4\right) \qquad \rho _{R}=a\,\theta ^{4}\text{ equations of state }%
\rho _{R}=\frac{1}{3}p\text{ }
\end{array}
\right.   \label{f11}
\end{equation}

\subsection{\textbf{Multiplicity of the dimensional basis.}}

From the above equations (\ref{f11}) we obtain five $\pi -$monomia. The
matrix of the exponents is: 
\begin{eqnarray}
&&
\begin{array}{l}
\pi _{6}:=\,t^{-2}G^{-1}c^{2}p^{-1} \\ 
\pi _{7}:=t^{-2}G^{-1}c^{2}\rho _{R}^{-1}\, \\ 
\pi _{8}:=\,\rho _{R}\,a^{-1}\,\theta ^{-4} \\ 
\pi _{9}:=\,\rho _{R}f^{4}A^{-1} \\ 
\pi _{10}:=\,\rho _{R}^{-1}p
\end{array}
\quad  \\
&&
\begin{array}{r|rrrrrrrrr}
& t & G & c & p & \rho _{R} & a & f & A & \theta  \\ \hline
\pi _{6} & -2 & -1 & 2 & 1 & 0 & 0 & 0 & 0 & 0 \\ 
\pi _{7} & -2 & -1 & 2 & 0 & 1 & 0 & 0 & 0 & 0 \\ 
\pi _{8} & 0 & 0 & 0 & 0 & 1 & -1 & 0 & 0 & -4 \\ 
\pi _{9} & 0 & 0 & 0 & 0 & 1 & 0 & 4 & -1 & 0 \\ 
\pi _{10} & 0 & 0 & 0 & 1 & -1 & 0 & 0 & 0 & 0
\end{array}
\end{eqnarray}
that leads us to a multiplicity of the base for this model of $4$. A
possible base is: $B=\left\{ f,\rho _{R},t,\theta \right\} \approx \left\{
L,M,T,\Theta \right\} ,$ where $\Theta $ stands for dimension of temperature.

\subsection{\textbf{Quantities and Constants.}}

According to what has been stated above, in section ($3.2$), we consider the
following set of quantities and constants in this model, written in the
dimensional base $B=\left\{ L,M,T,\Theta \right\} $

\begin{enumerate}
\item  \emph{$t,$ cosmic time (fundamental quantity) }$\left[ t\right] =T$

\item  \emph{$c,$ speed of light: }$\left[ c\right] =LT^{-1}$

\item  \emph{$G,$ gravitational constant: }$\left[ G\right]
=L^{3}M^{-1}T^{-2}$

\item  \emph{$a,$ radiation constant: }$\left[ a\right] =L^{-1}M^{1}T^{-2}%
\Theta ^{-4}$

\item  \emph{$A,$ Characteristic constant of the model }$\left[ A\right]
=L^{3}M^{1}T^{-2}$
\end{enumerate}

Therefore, by using the dimensional base $B$ the derived quantities will
appear in function of the unavoidable constants $(G,c,a,A)$ and the
fundamental quantity, $t.$

\subsection{\textbf{Solution through D.A.}}

We would like to obtain expressions for the temperature $\theta ,$ the
energy density $\rho _{R},$ the radius of the universe $f(t)$ (the latter
quantity is fundamental since it determines the metric and therefore the
geometry of our space-time) and finally the entropy $s$ and entropy density $%
S.\bigskip $

In this section we shall take into account the Barenblatt`s criterium (\cite
{B}) that eventually will allow solutions of the type $\pi _{i}=\left( \pi
_{j}\right) ^{n}.$

\subsubsection{\textbf{Calculation of the temperature.}}

We go next to carry out the calculation by three methods. The first one
through dimensional analysis i.e. by writing the matrix of the exponents and
applying the Pi theorem, a second one, also dimensional, by using the \emph{%
Planck`s system of units.} In both methods we must take into account the 
\textbf{Barenblatt`s criterium} to arrive to the complete solution of the
problem. We end showing an alternative way that brings us to a complete
solution of the equation avoiding in such a way the Barenblatt criterium,
since the use of numerical data is always is unsafe.\bigskip

We assume $\theta \propto \theta (G,c,A,a,t)$ ,where we designate for
dimensions of $\theta $ is $\left[ \theta \right] =\Theta $%
\begin{equation}
\begin{array}{r|rrrrrr}
& \theta & G & c & A & a & t \\ \hline
L & 0 & 3 & 1 & 3 & -1 & 0 \\ 
M & 0 & -1 & 0 & 1 & 1 & 0 \\ 
T & 0 & -2 & -1 & -2 & -2 & 1 \\ 
\Theta & 1 & 0 & 0 & 0 & -4 & 0
\end{array}
\label{m1}
\end{equation}
We obtain $2$ dimensionless $\pi -$monomia: 
\begin{equation}
\pi _{11}=\frac{\theta ca^{\frac{1}{4}}t}{A^{\frac{1}{4}}}\qquad \pi _{12}=%
\frac{GA}{c^{6}t^{2}}
\end{equation}
The solution that classic D.A. gives us is: 
\begin{equation}
\pi _{11}=\varphi (\pi _{12})\qquad \Longrightarrow \quad \theta =\frac{A^{%
\frac{1}{4}}}{ca^{\frac{1}{4}}t}\varphi \left( \frac{GA}{c^{6}t^{2}}\right)
\label{w1}
\end{equation}
where $\varphi $ represents an unknown function.\bigskip

If we take into account the Barenblatt`s criterium (\cite{B}), then we can
suppose that the solution is of the form $\pi _{11}=\left( \pi _{12}\right)
^{n}$. For this, we need to know the orders of magnitude of each one of the $%
\pi -$monomials  (\footnote{
See in section VI the table I of numerical values}): 
\begin{equation}
\pi _{11}=\frac{\theta _{0}ca^{\frac{1}{4}}t_{0}}{A^{\frac{1}{4}}}\approx
10^{2.614}\qquad \pi _{12}=\frac{GA}{c^{6}t_{0}^{2}}\approx 10^{-10.457}
\end{equation}
indicating us that the solution can be expressed as:

\begin{equation}
\pi _{11}=\left( \pi _{12}\right) ^{n}\qquad \Longrightarrow \qquad \theta
\propto \frac{A^{\frac{1}{4}}}{ca^{\frac{1}{4}}t}\left( \frac{GA}{c^{6}t^{2}}%
\right) ^{n}
\end{equation}
\begin{equation}
n=\frac{\log \pi _{11}}{\log \pi _{12}}=\frac{2.614}{-10.457}=-0.250\cong -%
\frac{1}{4}
\end{equation}
Then we have obtained through numerical calculation $n\simeq -\frac{1}{4}$.
The final result coincides with the theoretical one except for a numerical
factor. 
\begin{equation}
\theta (t)\propto \left( \frac{c^{2}}{Ga}\right) ^{\frac{1}{4}}\cdot t^{-%
\frac{1}{2}}\qquad \quad \theta (t)=\left( \frac{3c^{2}}{32\pi Ga}\right) ^{%
\frac{1}{4}}\cdot t^{-\frac{1}{2}}  \label{g1}
\end{equation}

Now we are going to use the \textbf{Planck`s system of units}, and
Barenblatt`s criterium. Since we know that all the quantities depend only on 
$t$ , then we can suppose that the temperature will be given by a
dimensionless product involving Planck`s temperature, Planck`s time and the
cosmic time. The solution is evidently: 
\begin{equation}
\theta (t)\propto \theta _{p}\cdot \varphi \left( \frac{t_{p}}{t}\right)
\end{equation}
If we take into account the Barenblatt`s criterium, we may suppose that the
solution will be of the form: 
\begin{equation}
\Longrightarrow \qquad \theta (t)\propto \theta _{p}\cdot \left( \frac{t_{p}%
}{t}\right) ^{n}
\end{equation}
since 
\begin{equation}
\pi _{13}=\left( \frac{\theta _{0}}{\theta _{p}}\right) \approx
10^{-31.715}\qquad \pi _{14}=\left( \frac{t_{p}}{t_{0}}\right) \approx
10^{-63.522}
\end{equation}
obtaining the value of $n$ through a simple numerical calculation.

\begin{equation}
n=\left( \frac{\log \pi _{13}}{\log \pi _{14}}\right) =\frac{-31.715}{-63.522%
}=0.499\approx \frac{1}{2}
\end{equation}
The solution that we obtain is therefore: 
\begin{equation}
\theta (t)\propto \theta _{p}\cdot \left( \frac{t_{p}}{t}\right) ^{\frac{1}{2%
}}  \label{p1}
\end{equation}
This is another dimensional solution$.$ Evidently, both expressions coincide
(compare (\ref{p1}) after simplifying it (\footnote{%
See table II in section 6}) with (\ref{g1}))\bigskip .

We can explore other possibilities. By examining the relation between the
constants. For example always in our formulas the quotient $\frac{G}{c^{2}}$
holds . For this reason we define a single new constant $B=\frac{G}{c^{2}}$
where $\left[ B\right] =LM^{-1}.$ Whit this new constant the set of
fundamental quantities results: $M=M(B,A,t)$ while the dimensional base
still holds. This new consideration brings us to obtain the following
matrix: 
\begin{equation}
\begin{array}{r|rrrrr}
& \theta  & B & A & a & t \\ \hline
L & 0 & 1 & 3 & -1 & 0 \\ 
M & 0 & -1 & 1 & 1 & 0 \\ 
T & 0 & 0 & -2 & -2 & 1 \\ 
\Theta  & 1 & 0 & 0 & -4 & 0
\end{array}
\end{equation}
obtaining a single monomial that brings us to the following solution 
\begin{equation}
\theta (t)\propto \left( \frac{1}{Bat^{2}}\right) ^{\frac{1}{4}}=\left( 
\frac{c^{2}}{Ga}\right) ^{\frac{1}{4}}t^{-\frac{1}{2}}
\end{equation}

This method allows to avoid Barenblatt%
\'{}%
s criterium, always very uncertain, since it depends of observations.
Evidently all this considerations could be made in the cases bellow.

\subsubsection{\textbf{Calculation of energy density.}}

$\rho _{R}\propto \rho (G,c,A,t)$ where $\left[ \rho _{R}\right]
=L^{-1}MT^{-2}\Theta ^{0}.$ The dimensional equation of this quantity can
not depend of the constant $a$ since the resting quantities and constants
are independent of \ temperature. We have them: 
\begin{equation}
\begin{array}{r|rrrrr}
& \rho _{R} & G & c & A & t \\ \hline
L & -1 & 3 & 1 & 3 & 0 \\ 
M & 1 & -1 & 0 & 1 & 0 \\ 
T & -2 & -2 & -1 & -2 & 1
\end{array}
\end{equation}
obtaining: 
\begin{equation}
\pi _{15}=\frac{A}{c^{4}t^{4}\rho _{R}}\qquad \pi _{12}=\frac{GA}{c^{6}t^{2}}
\end{equation}
that lead us to: 
\begin{equation}
\rho _{R}\propto \frac{A}{c^{4}t^{4}}\cdot \varphi \left( \frac{GA}{%
c^{6}t^{2}}\right) 
\end{equation}

In this case we can not apply the Baremblatt`s criterium, since the absolute
values of the orders of magnitude coincide: 
\begin{equation}
\pi _{15}=\frac{c^{4}t_{0}^{4}\rho _{R_{0}}}{A}\approx 10^{10.457}\qquad \pi
_{12}=\frac{GA}{c^{6}t_{0}^{2}}\approx 10^{-10.457}
\end{equation}
however if we insist on assuming that the solution will be of the form $\pi
_{15}=\left( \pi _{12}\right) ^{n}$. Obviously $n=-1$ being admittedly 
\begin{equation}
\rho _{R}\propto \frac{c^{2}}{Gt^{2}}\qquad \qquad \qquad \rho _{R}=\frac{%
3c^{2}}{32\pi Gt^{2}}  \label{d1}
\end{equation}

Using the Planck`s system: 
\begin{equation}
\pi _{16}=\left( \frac{\rho _{R_{0}}}{\rho _{p}}\right) \approx
10^{-127.045}\qquad \pi _{14}=\left( \frac{t_{p}}{t_{0}}\right) \approx
10^{-63.522}
\end{equation}
we see after comparing the two $\pi -$monomials that we can apply the
Barenblatt`s criterium and we suppose that the solution has the form: 
\begin{equation}
\rho _{R}(t)\propto \rho _{p}\cdot \left( \frac{t_{p}}{t}\right) ^{n}
\label{p2}
\end{equation}
\begin{equation}
n\approx \left( \frac{\log \pi _{16}}{\log \pi _{14}}\right) =\frac{-127.045%
}{-63.522}=1.99\approx 2
\end{equation}
whit $n=2.$ Both expressions (compare (\ref{p2}) with (\ref{d1})) coincide.

We obtain the same solution by another approach (as Dirac in his LNH)
comparing the $\pi -$monomials. Since their orders of magnitude coincide we
can write 
\begin{equation}
\pi _{15}^{\prime }=\frac{A}{c^{4}t_{0}^{4}\rho _{R_{0}}}\approx
10^{-10.457}\qquad \pi _{12}=\frac{GA}{c^{6}t_{0}^{2}}\approx 10^{-10.457}
\end{equation}
\begin{equation}
\pi _{15}^{\prime }=\frac{A}{c^{4}t^{4}\rho _{R}}=\frac{GA}{c^{6}t^{2}}=\pi
_{12}\Longrightarrow \rho _{R}\propto \frac{c^{2}}{Gt^{2}}
\end{equation}

We can avoid the use of Baremblatt criterium if \ as in the section above we
consider the trick of using the new constant $B.$ In this case, $\rho =\rho
(A,B,t)$ such approach brings us to the following solution: 
\begin{equation}
\rho \propto \frac{1}{Bt^{2}}\qquad \rho _{R}\propto \frac{c^{2}}{Gt^{2}}
\end{equation}
observing again that this tactic is correct.

\subsubsection{\textbf{Calculation of the radius of the universe}}

This quantity $f(t)$ depends on $(G,c,A,t),$where $\left[ f\right] =L.$
Therefore now : 
\begin{equation}
\begin{array}{r|rrrrr}
& f & G & c & A & t \\ \hline
L & 1 & 3 & 1 & 3 & 0 \\ 
M & 0 & -1 & 0 & 1 & 0 \\ 
T & 0 & -2 & -1 & -2 & 1
\end{array}
\end{equation}
and we obtain two dimensionless products. The solution is: 
\begin{equation}
\pi _{17}=\frac{f}{ct}\qquad \pi _{12}=\frac{GA}{c^{6}t^{2}}\qquad
\Longrightarrow f\propto ct\cdot \varphi \left( \frac{GA}{c^{6}t^{2}}\right) 
\label{m3}
\end{equation}
where the orders of magnitude are: 
\begin{equation}
\pi _{17}=\frac{f_{0}}{ct_{0}}\approx 10^{-2.614}\qquad \pi _{12}=\frac{GA}{%
c^{6}t_{0}^{2}}\approx 10^{-10.457}
\end{equation}
this situation coincides with the paragraph $4.3.2$ (Calculation of the
temperature) 
\begin{equation}
f\propto ct\cdot \left( \frac{GA}{c^{6}t^{2}}\right) ^{n}
\end{equation}
i.e. the Barenblatt`s criterium enables us to take a solution of the type: $%
\pi _{17}=\left( \pi _{12}\right) ^{n}$. Then we calculate $n$ as: 
\begin{equation}
\pi _{17}=\left( \pi _{12}\right) ^{n}\qquad \Rightarrow \quad n\approx
\left( \frac{\log \pi _{1}}{\log \pi _{2}}\right) \approx \frac{1}{4}
\end{equation}
and we can write the following expression: 
\begin{equation}
f(t)\propto \left( \frac{GA}{c^{2}}\right) ^{\frac{1}{4}}t^{\frac{1}{2}%
}\qquad \qquad f(t)=\left( \frac{32\pi GA}{3c^{2}}\right) ^{\frac{1}{4}}t^{%
\frac{1}{2}}  \label{s1}
\end{equation}
Also the orders of magnitude in absolute value of $\pi _{17}$ given by (\ref
{m3}) and $\pi _{11}$ from (\ref{m1}) are the same (this is related to the
hypothesis LNH\ of Dirac) proving that today: 
\begin{equation}
\pi _{17}^{\prime }=\frac{ct_{0}}{f_{0}}\approx 10^{2.614}\qquad \pi _{11}=%
\frac{\theta _{0}ca^{\frac{1}{4}}t_{0}}{A^{\frac{1}{4}}}\approx 10^{2.614}
\end{equation}
\begin{equation}
\frac{ct}{f}=\frac{\theta ca^{\frac{1}{4}}t}{A^{\frac{1}{4}}}\Longrightarrow
f=\frac{A^{\frac{1}{4}}}{\theta a^{\frac{1}{4}}}\qquad \Longleftrightarrow \
\rho _{R}f^{4}=A
\end{equation}
this solution verifies the equations (if we substitute $\theta a^{\frac{1}{4}%
}$ by its value calculated in equation (\ref{g1}) we obtain (\ref{s1}) as a
result).\bigskip 

The differential equation to be solved is now: 
\begin{equation}
f^{2}(f^{\prime })^{2}=\frac{8\pi G}{3c^{2}}A
\end{equation}
through D.A. we can integrate it easily obtaining the solution (\ref{s1}).
We observe that the governing parameters in this case are $M=M(t,G,c,A)$ and
if we do the same trick as before we can express the relation $\frac{G}{%
c^{2}}=B,$ obtaining a single $\pi -monomia$ avoiding the Baremblatt
criterium. We can increase the simplicity of the problem if we do $\frac{GA%
}{c^{2}}=N$ where $\left[ N\right] =L^{4}T^{-2\text{ }}$ and using a simple
dimensional base $B^{\prime }=\left\{ L,T\right\} .$ In this case the
solution is: 
\begin{equation}
f\propto N^{1/4}t^{1/2}
\end{equation}
obviously if we simplify $N$ for its valor it is observed that we recover
the classical solution 
\begin{equation}
f\propto N^{1/4}t^{1/2}\qquad f\propto \left( \frac{GA}{c^{2}}\right)
^{1/4}t^{1/2}
\end{equation}

In the calculation of this quantity we cannot use the Planck`s system of
units as before. This question is known as the Planck problem and it was
Zeldovich who pointed out this conflict in the standard model. Zeldovich
(see \cite{Z}) emphasized that this is perhaps the most fundamental and
serious problem of the standard cosmology. This mismatch of scales is
generally referred to as the Planck problem.

\subsubsection{\textbf{Calculation of the Entropy.}}

Equations (\ref{f11}) correspond to no variation of entropy in the Universe.
We mean to calculate the entropy of this Universe, $s,$ and the entropy
density, $S$.\bigskip

For its calculation we follow the same method, but in this case we do not
consider $t$ since we know that this quantity is constant (\cite{N})
consequently:

$s\propto s(G,c,A,a)$ where $\left[ s\right] =L^{2}MT^{-2}\Theta ^{-1}$ and
we get: 
\begin{equation}
\begin{array}{r|rrrrr}
& s & G & c & A & a \\ \hline
L & 2 & 3 & 1 & 3 & -1 \\ 
M & 1 & -1 & 0 & 1 & 1 \\ 
T & -2 & -2 & -1 & -2 & -2 \\ 
\Theta  & -1 & 0 & 0 & 0 & -4
\end{array}
\end{equation}
\begin{equation}
\Longrightarrow s\propto \left( A^{3}a\right) ^{\frac{1}{4}}
\end{equation}
This value is too high, being a difficult issue of justifying within the
model.\bigskip 

We suppose that we ignore the behavior of this quantity (we do not know that 
$s=const$.). In this case we could follow up the exposed method up to now,
that is to say, we would calculate this quantity in function of $%
(t,G,c,a,A). $ In this case the utilization of the constant $a$ is necessary
for dimensional considerations. As we have seen, we obtained two monomials: $%
\pi $$_{18}=\left( \frac{s}{\left( A^{3}a\right) ^{\frac{1}{4}}}\right) $
and $\pi _{12}=\left( \frac{GA}{c^{6}t^{2}}\right) .$ Then the solution that
we would obtain would be. $\pi _{18}=\varphi \left( \pi _{12}\right) .$
Taking into account the criterion of Baremblatt and knowing that: $s\approx
10^{64.18735}JK^{-1}$ we get, $\pi _{18}\approx 10^{0},$ this enables us to
write $\pi _{18}=\left( \pi _{12}\right) ^{n}$ iif  $n=0$ obtaining therefore
the solution that gives us our first position.\bigskip

As above, we can explore the possibility of reducing the number of
constants. In this case such reduction brings us to the following solution: 
\begin{equation}
s=s(A,B,t)\qquad \Longrightarrow s\propto \left( A^{3}a\right) ^{\frac{1}{4}}
\end{equation}

For the calculation of $S\propto S(G,c,A,a,t)$%
\begin{equation}
\begin{array}{r|rrrrrr}
& S & G & c & A & a & t \\ \hline
L & -1 & 3 & 1 & 3 & -1 & 0 \\ 
M & 1 & -1 & 0 & 1 & 1 & 0 \\ 
T & -2 & -2 & -1 & -2 & -2 & 1 \\ 
\Theta  & -1 & 0 & 0 & 0 & -4 & 0
\end{array}
\end{equation}
\begin{equation}
S\propto \left( \frac{a^{\frac{1}{4}}A^{\frac{3}{4}}}{c^{3}t^{3}}\right)
\cdot \varphi \left( \frac{GA}{c^{6}t^{2}}\right) 
\end{equation}
as we do not know here the numerical values of such quantities we can not
operate as before. But we know that $S=s/f^{3}.$ Simplifying both
expressions and eliminating $a$ with $\left( a\propto \frac{k_{B}}{%
c^{3}\hbar ^{3}}\right) $ we obtain 
\begin{equation}
S\propto \left( \frac{a^{\frac{1}{4}}c^{\frac{6}{4}}}{G^{\frac{3}{4}}t^{%
\frac{3}{2}}}\right) \propto \left( \frac{ac^{6}}{G^{3}t^{6}}\right) ^{\frac{%
1}{4}}\propto k_{B}\left( \frac{c}{G\hbar t^{2}}\right) ^{\frac{3}{4}}
\label{t1}
\end{equation}
Observe that $S\propto a\theta ^{3}.$ If we simplify this expression then we
obtain the results above (\ref{t1}). If we reduce the number of constants,
this tactic brings us to obtain a single monomial: 
\begin{equation}
S\propto \left( \frac{a}{B^{3}t^{6}}\right) ^{\frac{1}{4}}\qquad
\Longrightarrow S\propto \left( \frac{ac^{6}}{G^{3}t^{6}}\right) ^{\frac{1}{4%
}}
\end{equation}

\section{\textbf{Conclusions.}}

We have shown formally the fruitful application of the D.A. to these two
cosmological concrete models. For the case of the Einstein-de Sitter one, we
have arrived at the solution in a trivial way since it appears a single
dimensionless product, whereas for the model type FRW with radiation
predominance we have explored various possibilities in particular we have
taken into account the criterion of Barenblatt in order to obtain the
complete solution of the equations. For the latter model we have justified,
from a dimensional point of view, the utilization of the Planck`s system of
units. We believe therefore that the method developed here can be useful for
solving more complex models whose equations might be of difficult
integration. We try to be rigorous when formalizing all the required steps.
But if one has got a good knowledge (from the physical point of view) of the
behavior of the model one does not need be so scrupulous. There is no need
of developing the equations to obtain relationships between the quantities
that form part of the model. We think therefore that this method may also
have some pedagogical interest.

\section{\textbf{Table of quantities and constants.}}

\begin{center}
\begin{tabular}{|l|l|}
\hline
\textbf{Quantity I.S.} & \textbf{Constant I.S.} \\ \hline\hline
$\theta _{0}\approx 10^{0.436}$$K$ & $G\approx 10^{-10.1757}$$%
m^{3}kg^{-1}s^{-2}$ \\ \hline
$f_{0}\approx 10^{26}$$m$ & $c\approx 10^{8.476821}$$ms^{-1}$ \\ \hline
$t_{0}\approx 10^{20.252}$$s$ & $a\approx 10^{-15.121153}$$Jm^{-3}K^{-4}$ \\ 
\hline
$\rho _{R_{0}}\approx 10^{-13.379}Jm^{-3}$ & $A\approx 10^{90.62}$ $%
m^{3}kgs^{-2}$ \\ \hline
\end{tabular}
\end{center}

Where $\theta _{0}$ represents the temperature of background cosmic
microwave radiation today i.e. $\theta _{0}\approx 2.73K\Rightarrow \log
{}_{10}(2.73)=0.436162$ $\Rightarrow \theta _{0}\approx 10^{0.436}K$ and $%
f_{0}$ is the radius of the Universe, $t_{0}$ represents the approximate age
of the Universe and $\rho _{R_{0}}$ is the energy density of the radiation
today

\underline{\textbf{Planck system:}} length, time, mass, energy density and
temperature.

\begin{center}
\begin{tabular}{|c|c|c|}
\hline
\textbf{Quantity} & \textbf{Definition} & \textbf{N. value I.S.} \\ 
\hline\hline
$l_{p}$ & $\sqrt{\frac{G\hbar }{c^{3}}}$ & $10^{-34.7915}$$m$ \\ \hline
$t_{p}$ & $\sqrt{\frac{G\hbar }{c^{5}}}$ & $10^{-43.2684}$$s$ \\ \hline
$m_{p}$ & $\sqrt{\frac{\hbar c}{G}}$ & $10^{-7.6622}$ $kg$ \\ \hline
$\rho _{p}$ & $\frac{m_{p}c^{2}}{l_{p}^{3}}$ & $10^{113.666}Jm^{-3}$ \\ 
\hline
$\theta _{p}$ & $\sqrt{\frac{\hbar c^{5}}{k_{B}^{2}G}}$ & $10^{32.1514}$$K$
\\ \hline
\end{tabular}
\end{center}

\textbf{{\LARGE A}CKNOWLEDGMENTS}

I wish to thank Prof. Casta\~{n}s for suggestions and enlightening
discussions and to Javier Aceves for helping in the translation into English.\\


\bigskip

\begin{thebibliography}{9}
\bibitem{N}  \textbf{Narlikar, J.V.} Introduction to Cosmology. CUP 1993.

\bibitem{T}  \textbf{Belinch\'{o}n, J. A}.Gen. Rel.
and Grav (2000) \textbf{32}, 1487-98. \textbf{Belinch\'{o}n, J. A}. acepted in Class.
Quant. Grav (2000) \textbf{17}. \textbf{Belinch\'{o}n, J. A}. gr-qc/9907003 acepted in Int. Jour. Theor. Phys. 

\bibitem{P}  \textbf{Palacios, J. }Dimensional Analysis. Macmillan 1964
London.

\bibitem{C}  \textbf{Casta\~{n}s, M.} Sobre el primer postulado de Palacios.
JTGAD XXXI 1995.Preprint Group of D.A. Dept. of Physic ETS Architecture UP
Madrid.

\bibitem{B}  \textbf{Barenblatt, G.I. }Scaling, self-similarity and
intermediate asymptotics. Cambridge texts in applied mathematics N 14 1996
CUP.

\bibitem{K}  \textbf{Kurth, R}. Dimensional Analysis and Group Theory in
Astrophysics. Pergamon 1972

\bibitem{L}  \textbf{Falla, D. F.\&Landsberg, P}. Black holes and limits on
some physical quantities. EJP, \textbf{15}, 204-211, (1994)

\bibitem{Z}  \textbf{Zeldovich, Ya.B.} My Universe. Selected reviews. 1992
pp 95 Harwood Acd. Press
\end{thebibliography}
\end{document}